\documentclass[a4paper,11pt,twoside,rotating]{article}
\usepackage{epsfig,psfrag,rotating}
\newcommand{\imag}{\Im {\rm m}}
\newcommand{\real}{\Re {\rm e}}
\def\mueg{\mu^- \to e^- \gamma}
\def\taueg{\tau^- \to e^- \gamma}

\newcommand {\misset}  {E_T\hspace{-4mm}/ \hspace{1mm}}

\def\ti{\tilde}
\newcommand{\T}{\tilde\tau}
\newcommand{\SN}{\tilde\nu_{\tau}}

\newcommand{\lsim}{\;\raisebox{-0.9ex}{$\textstyle\stackrel{\textstyle <}
           {\sim}$}\;}

\def\b   {\beta}
\def\t   {\theta}

\def\tsa {\theta_{\sa}}

\def\st  {\ti t}

\def\sa  {\ti \tau}

\def\beq{\begin{equation}} 
\def\eeq{\end{equation}} 
\def\bea{\begin{eqnarray}}  
\def\eea{\end{eqnarray}}  
\def\beano{\begin{eqnarray*}}  
\def\eeano{\end{eqnarray*}}  
\def\half{{{\textstyle{1 \over 2}}}}  
\def\noin{\noindent}

\def\tm{{\tilde \mu}}
\def\st{{\tilde \tau}}

\begin{document}
\begin{flushright}
IFIC-02-67\\
IFT- 02/44 \\
UWThPh-2002-39\\
ZU-TH 20/02\\
hep--ph/0301027\\
\end{flushright}
\vskip1.5cm
\begin{center}
{\LARGE\bf CP Phases, LFV,  RpV \& all that}\\[1cm]

{
 A.~Bartl$^1$, K.~Hidaka$^2$, M.~Hirsch$^3$,
  J.~Kalinowski$^4$\footnote{Convener and Rapporteur of the ECFA/DESY
  SUSY Collaboration. Talk at the 
International Workshop on Linear Colliders LCWS(2002), August 26-30,
2002, Jeju, Korea.},
T.~Kernreiter$^3$,\\
W.~Majerotto$^5$,  W.~Porod$^6$, J.C.~Rom\~ao$^7$, J.W.F.~Valle$^3$\\[2ex]
{\it ECFA/DESY SUSY Collaboration}\\[2ex]
{\it $^1$ Inst.\ f\"ur Theoretische Physik, Universit\"at Wien, A-1090
  Vienna, Austria}\\
{\it $^2$ Dept.\ Physics, Tokyo Gakugei University, Koganei, Tokyo,
  184-8501, Japan}\\
{\it $^3$ Instituto de F\'isica Corpuscular, E-46071 Valencia, Espa\~na}\\
{\it $^4$ Inst.\ of Theoretical Physics, Warsaw University, 00681
  Warsaw, Poland}\\
{\it $^5$ Inst.\ f\"ur  Hohenenergiephysik, \"OAW, A-1050 Vienna,
  Austria}\\ 
{\it $^6$ Inst.\  Theor.\ Physik, Universit\"at Z\"urich, CH-8057 Z\"urich,
  Switzerland}\\
{\it $^7$ Dept.\ de Fisica, Instituto Superior T\'ecnico, 1049-001 Lisboa,
  Portugal}
}

\vskip2cm
\thispagestyle{empty}

{\bf Abstract}\\[1pc]

\begin{minipage}{13cm}
Most phenomenological analyses of searches for supersymmetric
particles  have been
performed within the MSSM with real SUSY parameters and conserved
R-parity and lepton flavour. Here we summarize recent results obtained
in the (s)lepton sector when one of the above assumptions is relaxed.    
\end{minipage}  \\
\end{center}

\vfill
\newpage

\setcounter{page}{1}

\title{CP Phases, LFV,  RpV \& all that }

\author{A.~Bartl$^1$, K.~Hidaka$^2$, M.~Hirsch$^3$,
  J.~Kalinowski$^4$\footnote{Convener and Rapporteur of the ECFA/DESY
  SUSY Collaboration},
T.~Kernreiter$^3$,\\
W.~Majerotto$^5$,  W.~Porod$^6$, J.C.~Rom\~ao$^7$, J.W.F.~Valle$^3$\\[2ex]
{\it ECFA/DESY SUSY Collaboration}\\[2ex]
{\it $^1$ Inst.\ f\"ur Theoretische Physik, Universit\"at Wien, A-1090
  Vienna, Austria}\\
{\it $^2$ Dept.\ Physics, Tokyo Gakugei University, Koganei, Tokyo,
  184-8501, Japan}\\
{\it $^3$ Instituto de F\'isica Corpuscular, E-46071 Valencia, Espa\~na}\\
{\it $^4$ Inst.\ of Theoretical Physics, Warsaw University, 00681
  Warsaw, Poland}\\
{\it $^5$ Inst.\ f\"ur  Hohenenergiephysik, \"OAW, A-1050 Vienna,
  Austria}\\ 
{\it $^6$ Inst.\  Theor.\ Physik, Universit\"at Z\"urich, CH-8057 Z\"urich,
  Switzerland}\\
{\it $^7$ Dept.\ de Fisica, Instituto Superior T\'ecnico, 1049-001 Lisboa,
  Portugal}
}
\date{}
\maketitle
\begin{abstract}
Most phenomenological analyses of searches for supersymmetric
particles  have been
performed within the MSSM with real SUSY parameters and conserved
R-parity and lepton flavour. Here we summarize recent results obtained
in the (s)lepton sector when one of the above assumptions is relaxed.    
\end{abstract}

Since supersymmetry must be broken at low energy, and the mechanism of
its breaking is still unknown, even the minimal supersymmetric model
(MSSM) introduces more than 100 new parameters.  The MSSM is
understood as an effective low energy model defined by a)~minimal
particle content, b)~$R$-parity conservation, c)~most general soft
supersymmetry breaking terms.  The number of parameters can be further
enlarged by relaxing a) or b), or reduced by constraining c) with
additional assumptions on SUSY breaking parameters. So far most
phenomenological studies on supersymmetric particle searches have been
performed within the MSSM with drastically reduced number of
parameters by assuming e.g. that all SUSY parameters are
real, lepton flavour is conserved, universality at high scale holds 
etc. However, current experimental limits on the SUSY parameter space
admit many of the above assumptions to be relaxed. We briefly present
some phenomenological consequences in the (s)lepton sector of i)
complex phases, ii) lepton flavour violation, iii) R-parity violation.
\section{CP phases}
The assumption of real SUSY parameters has partly been justified
by the experimental limits on the electric
dipole moments (EDM) of the electron, neutron and mercury atom. However, 
the EDM constraints can be avoided assuming 
masses of the first and second generation sfermions large
(above the TeV scale), or arranging 
cancellations between the 
different SUSY contributions to the EDMs \cite{nath}. As a result, 
the complex phase of the Higgsino mass parameter $\mu$ 
is much less restricted than previously assumed, whereas the
complex phases of the soft--breaking trilinear scalar 
coupling parameters $A_f$ are practically
unconstrained.

Recently  an analysis of production and decay rates of 
$\tilde \tau_1$, $\tilde \tau_2$ and $\tilde \nu_{\tau}$
at an $e^+e^-$ 
linear collider with a CMS energy $\sqrt{s}=0.5-1.2$~TeV 
with complex $\mu$, $A_{\tau}$ and $M_1$  
($M_1$ is the $U(1)$ gaugino mass
parameter) has been performed \cite{bhkp}. 
Explicit $CP$ violation in the Higgs sector
induced by stop and sbottom loops with complex parameters
\cite{carenaetal} has also been included, and  
the scalar mass matrices and trilinear
scalar coupling parameters have been taken flavor diagonal.

The stau mass matrix in the interaction 
basis  $(\tilde \tau_L^{},\tilde \tau_R^{})$ reads: 
\begin{eqnarray}   
{\cal M}^2_{\tilde\tau} &=& 
     \left( \begin{array}{cc} 
                m_{\sa_L}^2 & a_\tau^* m_\tau \\
                a_\tau m_\tau     & m_{\sa_R}^2
            \end{array} \right),\\ 
  m_{\sa_L}^2 &=& M_{\ti L}^2 
                  + m_Z^2\cos 2\beta\,(\sin^2\t_W - \half) 
                  + m_\tau^2,    \\
  m_{\sa_R}^2 &=& M_{\ti E}^2  
                  - m_Z^2 \cos 2\b\, \sin^2\t_W + m_\tau^2,  \\
  a_\tau m_\tau     &=& 
                     ({ A_\tau} - { \mu}^* \tan\beta)\,m_\tau
     \,\, 
                  = \,\, |a_\tau m_\tau| \, e^{i\varphi_{\sa}}. 
\end{eqnarray}
where $M_{\ti L,\ti E}$ and $A_\tau$ are slepton soft SUSY--breaking 
parameters, with $A_\tau = |A_\tau| \,
e^{i{ \varphi_{A_\tau}}}$  
and $\mu = |\mu| \, e^{i{ \varphi_{\mu}}}$. 
The  mass eigenstates are defined as 
\begin{eqnarray}
      \left( \begin{array}{c} 
                \sa_1 \\
                \sa_2
            \end{array} \right)& =& 
     \left( \begin{array}{cc} 
                e^{i\varphi_{\sa}} \cos\tsa & \sin\tsa \\
                -\sin\tsa  & e^{-i\varphi_{\sa}} \cos\tsa
            \end{array} \right) 
     \left( \begin{array}{c} 
                \sa_L \\
                \sa_R
            \end{array} \right).
\end{eqnarray}
In principle, the imaginary parts of the complex parameters involved
could most directly and unambiguously be determined by measuring
suitable $CP$ violating observables. However, in the $\ti
\tau_i$-system this is not straightforward, because the $\ti \tau_i$
are spinless and their main decay modes are two--body decays. On the
other hand, also the $CP$ conserving observables depend on the phases
of the underlying complex parameters, because the mass eigenvalues and
the couplings involved are functions of these parameters.

The masses  $m_{\sa_{1,2}}^2$ and mixing angle 
$\tsa$ depend on the phases only through a
term { $m_\tau^2|A_\tau\mu|\tan\beta\cos(\varphi_{A_\tau} + \varphi_{\mu})$}
\cite{bhkp}. Therefore $m_{\sa_{1,2}}^2$  
are essentially independent of the phases because $m_\tau$ is
small, whereas the phase dependence of $\tsa$ is strongest if { $|A_\tau|
\simeq |\mu| \tan\b$} and $|m_{\sa_L}^2 - m_{\sa_R}^2| \lsim |a_\tau
m_\tau|$.  Since the  $Z\sa_i\sa_i$ couplings are real, the $\sa_i\bar{\sa}_j$
production cross
sections do not explicitly depend on the phases (although
$Z\sa_1\sa_2$ coupling is complex, for $\sa_1\bar{\sa}_2$ production 
only $Z$ exchange
contributes). However, the various $\tilde\tau$ 
decay branching ratios depend in a
characteristic way on the complex
phases. This is illustrated in Fig.~1. 
\begin{figure}[t]
\epsfig{figure=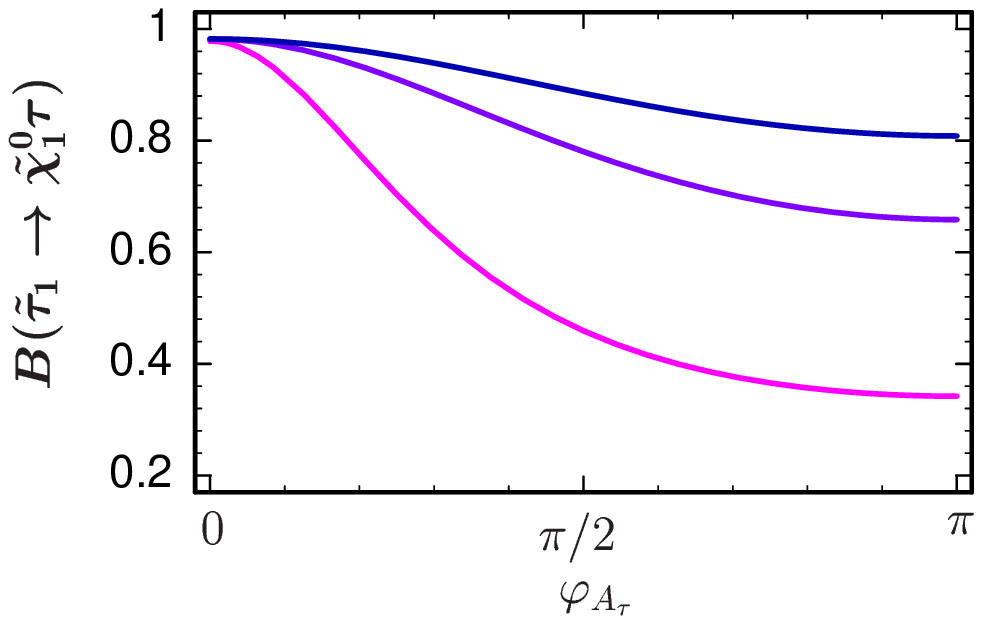, width=4.2cm,height=5cm}
\hspace{2mm}
\epsfig{figure=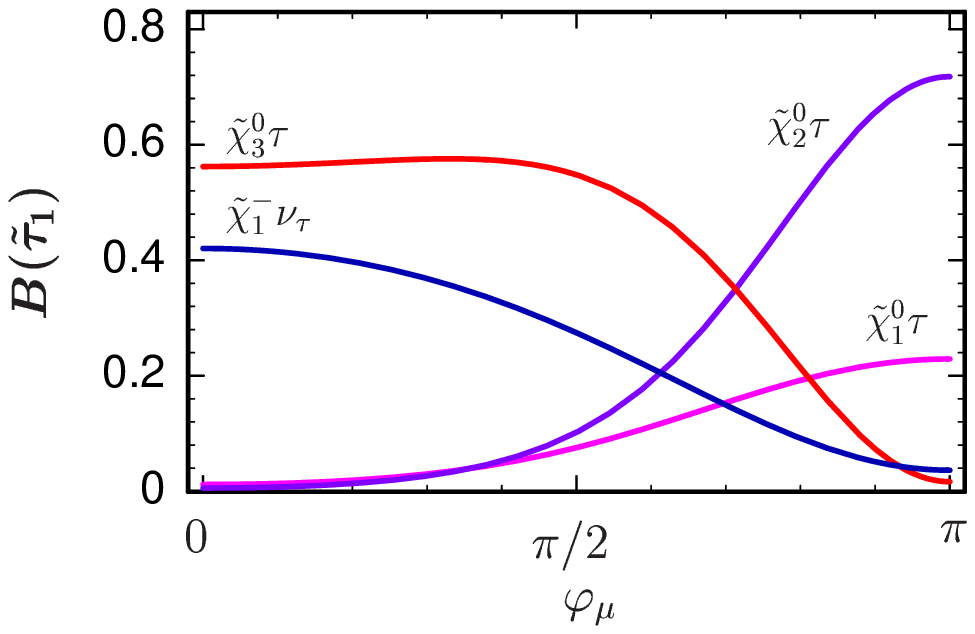, width=4.2cm,height=5cm}
\hspace{2mm}
\epsfig{figure=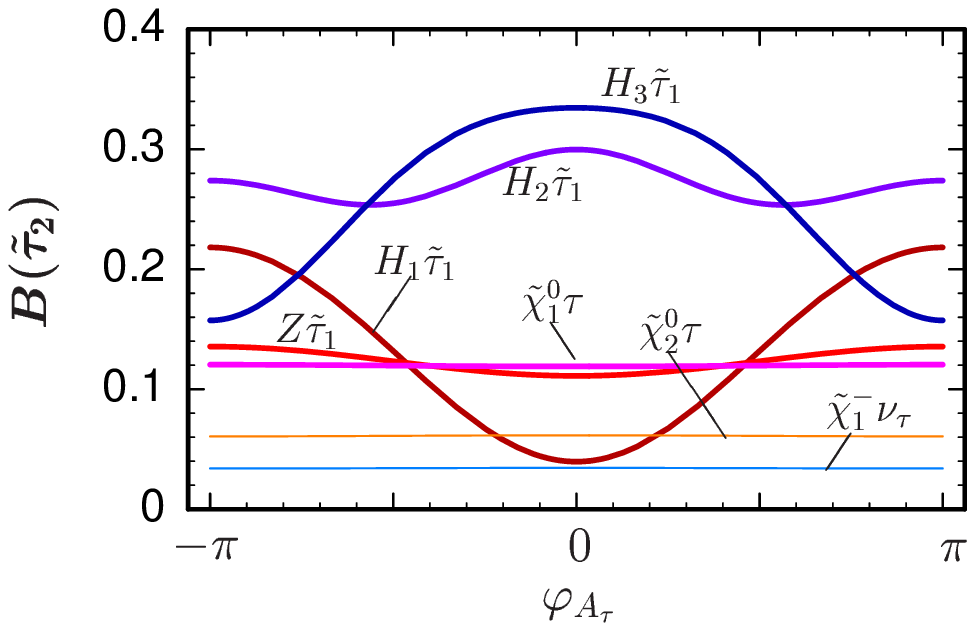, width=4.2cm,height=5cm}
\label{fig:taubr}
\vspace{-2mm}
\caption{Branching ratios of $\T_1$ and $\T_2$ as indicated. Left: for 
$m_{\T_1}=240$, $m_{\SN}=233$, 
$238$, $243$ (from bottom to top),
and $\varphi_{\mu} = \varphi_{U(1)} = 0$,
$|\mu| = 300$, $|A_{\tau}| = 1000$, $\tan\beta = 3$, and
$M_2 = 200$. Center: for 
$\varphi_{U(1)} = \varphi_{A_{\tau}} = 0$, 
$m_{\T_1}=240$, $m_{\T_2}=500$, 
$M_2=280$, $|\mu|=150$, $\tan\beta=3$, and $|A_{\tau}| =
1000$, assuming $M_{\ti L} < M_{\ti E}$. Right: for $\varphi_{\mu}=0$ 
$m_{\T_1}=240$, $m_{\T_2}=500$,
$m_{H^{\pm}}=160$, $|\mu|=600$, $M_2=450$, 
$\varphi_{U(1)}=0$, $\tan\beta=30$, and $|A_{\tau}| = 900$,
assuming $M_{\ti L} > M_{\ti E}$. All mass parameters are in GeV.}
\end{figure}
The fit to the simulated experimental
 data with 2 ab$^{-1}$ at a collider like TESLA shows that  
 $\imag A_\tau$ and $\real
A_\tau$ can be determined with an error of order 10\%. 

\section{Lepton flavour violation}
Neutrino oscillation experiments have established the existence
of lepton flavour violation (LFV) with $\tan^2\theta_{Atm}\simeq 1$, 
$\tan^2\theta_{\odot} = 0.24-0.89$ and $\sin^2(2\theta_{13}) \lsim 0.1$  
\cite{nuosc}.
On the other hand,
there are stringent constraints on LFV in the charged
lepton sector, the strongest being 
$BR(\mueg) < 1.2 \times 10^{-11}$ \cite{Groom:2000in}. 

In supersymmetric models  gauge and Lorentz invariance does not
enforce 
total lepton number $L=L_e + L_\mu + L_\tau$ or individual lepton flavour
$L_e$,  $L_\mu$ or  $L_\tau$ to be conserved. 
One usually invokes R-parity symmetry, which
forces total lepton number conservation but still allows the violation of
individual lepton number, e.g.~due to loop effects in $\mueg$ \cite{bm}.
Moreover, in the MSSM a large $\nu_\mu$-$\nu_\tau$ mixing can lead to a 
large  $\tilde \nu_\mu$-$\tilde \nu_\tau$ mixing
via renormalisation group equations. Therefore one can
expect  clear LFV
signals in slepton and sneutrino production
and in the decays of neutralinos and charginos into sleptons and sneutrinos
at the LHC and at future lepton colliders. 

In ref.~\cite{ego} the consequences of LFV 
assuming the {\it most general} mass matrices for sleptons
and sneutrinos have been studied.
The charged slepton 
mass matrix, generalized to include flavour mixing as well as 
left-right mixing, is given by:
\begin{equation}
  M^2_{\tilde l} = \left(
    \begin{array}{cc}
      M^2_{L,ij} + \frac{1 }{2} v^2_d Y^{E*}_{ki} Y^{E}_{kj}
       + D_L  \delta_{ij}  &
        \frac{1}{\sqrt{2}} (v_d A_{ji} - \mu^* v_u Y^{E*}_{ij} )   \\
     \frac{1 }{\sqrt{2}} (v_d A^*_{ji} - \mu v_u Y^{E}_{ij}) & 
        M^2_{R,ij} + \frac{1}{2} v^2_d Y^{E}_{ik} Y^{E*}_{jk} 
      -  D_R \delta_{ij}   \\
     \end{array} \right), 
  \label{eq:sleptonmass}
\end{equation}
with $D_L =\frac{1}{8}( {g'}^2 -  g^2 ) (v^2_d - v^2_u)$ and
$D_R=\frac{1}{4} {g'}^2  (v^2_d - v^2_u)$, and 
the indices $i,j,k=1,2,3$ counting flavors $e,\mu,\tau$.
$M^2_{L}$ and $M^2_{R}$ are the soft SUSY breaking mass matrices for
left and right sleptons, respectively. $A_{ij}$ are the trilinear soft
SUSY breaking couplings of the sleptons and Higgs bosons, and
$Y^E_{ij}$  are charged lepton Yukawa couplings.
Similarly, one finds for the sneutrinos
\begin{equation}
  M^2_{\tilde \nu,ij} =  M^2_{L,ij} \textstyle
  + \frac{1}{8} \left( g^2 + {g'}^2 \right) (v^2_d - v^2_u) \delta_{ij}.
  \label{eq:sneutrinomass}
\end{equation}

For the numerical analysis the SPS1 reference point  
\cite{Georg} (defined by   $M_{1/2} = 250$~GeV,
$M_0=100$~GeV, $A_0=-100$~GeV, $\tan \beta = 10$ and sign$(\mu)=+$ at
the GUT scale) has been chosen, with
the following slepton mass parameters at the electroweak scale: 
$M_{R_{11}}$ = 138.7~GeV,
$M_{R_{33}}$ = 136.3~GeV, $M_{L_{11}}$ = 202.3~GeV, $M_{L_{33}}$ = 201.5~GeV
and $A_{33}/Y^E_{33}  = -257.3$~GeV. 
With these parameters fixed, 
a scan over the nondiagonal entries of  $M^2_L$, $M^2_R$ and $A$  shows that 
values for  $|M^2_{R,ij}|$ up to $8 \cdot 10^3$~GeV$^2$, $|M^2_{L,ij}|$
up to $6 \cdot 10^3$~GeV$^2$ and $|A_{ij} v_d|$ up to 650~GeV$^2$ 
are compatible 
with the current experimental 
constraints. In most cases, one of the mass squared parameters is at least
one order of magnitude larger than all the others. However, there is a
sizable part in parameters where at least two of the off-diagonal parameters
have the same order of magnitude.
 
\begin{figure}
\setlength{\unitlength}{1mm}
\begin{picture}(150,45)
\put(0,0){\mbox{\epsfig{figure=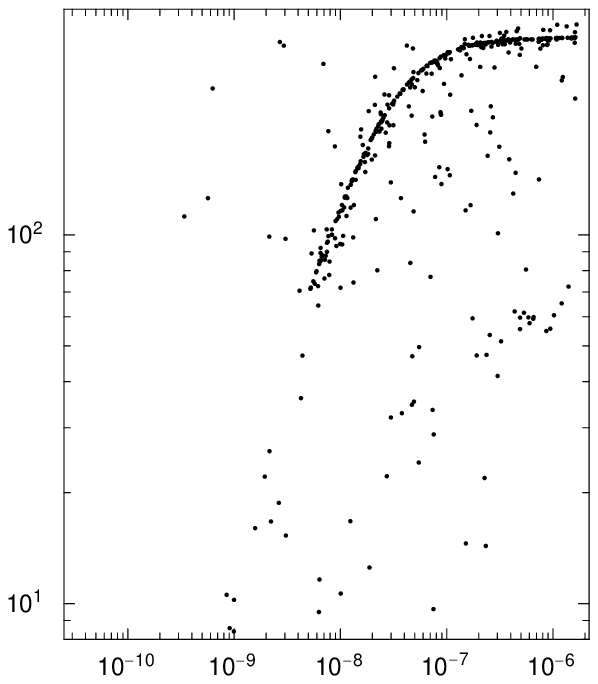,height=4cm,width=5.5cm}}}
\put(70,0){\mbox{\epsfig{figure=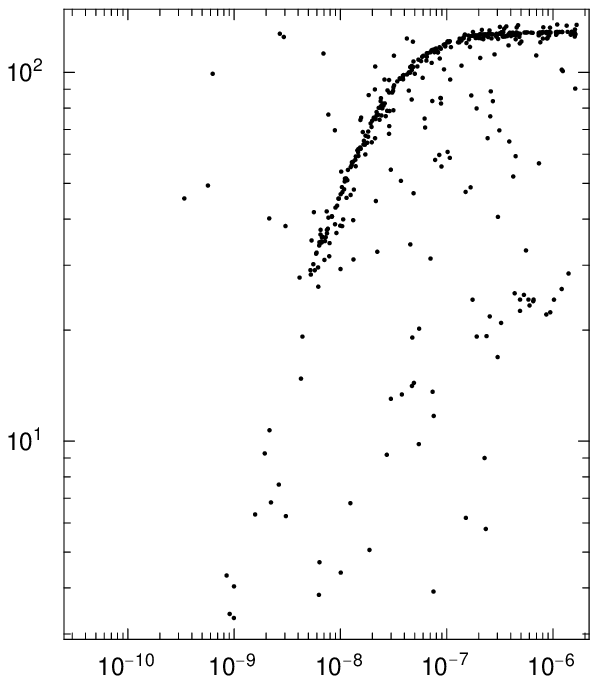,height=4cm,width=5.5cm}}}
\put(0,42){\mbox{\bf (a)}}
\put(8,42){\mbox{$\sigma(e^+ e^- \to
e^\pm \tau^\mp \misset)$~[fb]}}
\put(35,-3){\mbox{BR$(\tau \to e \gamma)$}}
\put(69,42){\mbox{\bf (b)}}
\put(77,42){\mbox{$signal/\sqrt{background}$}}
\put(105,-3){\mbox{BR$(\tau \to e \gamma)$}}
\end{picture}
\caption{({\bf a}) Cross section in fb for the signal $e^\pm \tau^\mp \misset$ 
            and  ({\bf b}) the ratio
           signal over square root of background  as a function of 
           BR$(\tau \to e \gamma)$ 
           for $\sqrt{s} = 500$~GeV, $P_{e^-} = 0$ and $P_{e^+} = 0$. In the
           latter case we have assumed an integrated luminosity of 
           100 fb$^{-1}$.}
\label{fig:signal1}
\end{figure}

Possible LFV signals at an 
$e^+ e^-$ collider include  $e \mu \,  \misset$, 
$e \tau \, \misset$, $\mu \tau \,  \misset$ in the final state 
plus a  
possibility of additional jets.  Varying
the parameters randomly on a logarithmic scale: $ 10^{-8} \le |A_{ij}|
\le 50$~GeV, $ 10^{-8} \le M^2_{ij} \le 10^4$~GeV$^2$,  8000
points consistent with the experimental data have been generated. 
In Fig.~\ref{fig:signal1} the cross section of $e^+ e^- \to
e^\pm \tau^\mp \misset$  and 
the corresponding ratio signal over square root of the background
($S/\sqrt{B}$) are shown as a function $BR(\taueg)$ assuming an integrated
luminosity of 100~fb$^{-1}$ at $\sqrt{s}=500$ GeV.  
All possible SUSY
 and Higgs cascade decays have been included together with  ISR- and
SUSY-QCD corrections for the production cross sections, while 
the background comes from  all possible SUSY cascade
decays faking the signal and the SM  
$W$, $t$-quark and $\tau$-lepton pair production processes. 
Although no cuts have been applied,
there is in most cases a spectacular signal. 
The accumulation of points in Fig.~\ref{fig:signal1} along a band is due to
a large $\tilde e_R$-$\tilde \tau_R$
mixing which is less constraint by $\taueg$ than the corresponding
left-left or left-right mixing. 

Note that the collider LFV signals can be very competitive to those
from rare charged lepton decay, like $\tau\to \mu \gamma$. This is 
illustrated in Fig.~3 
\cite{jk}, where for simplicity  the flavour mixing  has been 
restricted to the 2-3 generation subspace of sneutrinos with  the mixing angle
$\theta_{23}$ and $\Delta m_{23} = 
|{m}_{\tilde\nu_2} - {m}_{\tilde\nu_3}|$ as free, independent
parameters.  

\begin{figure}[h!]
\epsfig{figure=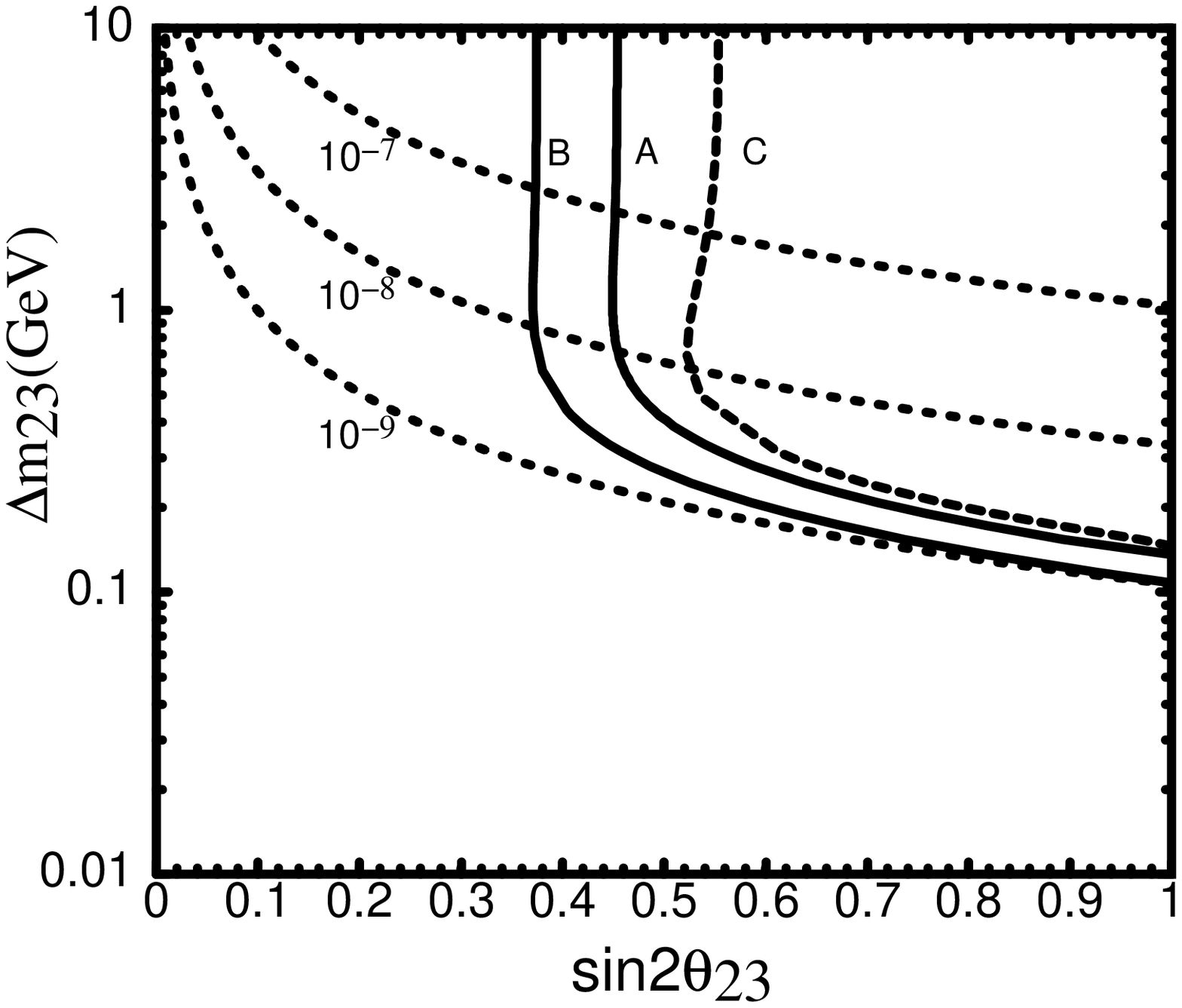,width=5.5cm,height=5cm}
\label{fig:lcbr}
\put(20,60){\mbox{\begin{minipage}{7cm}{Figure 3: 
3$\sigma$ significance contours in the $\theta_{23}$ and $\Delta
m_{23}$ plane for
 $\sqrt{s}=$500 GeV LC  and for luminosity 
 500 fb$^{-1}$ (A), and 1000 fb$^{-1}$ (B). 
 Line C: $\tilde\nu \tilde\nu^*$
 contribution with luminosity 500 fb$^{-1}$.  Dotted 
 lines: BR($\tau\to\mu\gamma$)=10$^{-7}$, 10$^{-8}$, 
 10$^{-9}$.}\end{minipage}}}
\end{figure}

\section{R-parity violation}
Supersymmetric models with explicit bilinear breaking of R-parity
(RpV) \cite{epsrad} provide a simple and calculable
framework for neutrino masses and mixing angles in agreement with the
experimental data. 
The simplest bilinear RpV model, studied in ref.~\cite{hprv},  is
characterized by three additional  terms $\epsilon_i$ in the superpotential
\begin{equation}
\label{eq:Wpot} 
W = W_{MSSM} + \epsilon_i \widehat
  L_i\widehat H_u,
\end{equation}  
and the corresponding terms in the
soft SUSY breaking part of the Lagrangian, 
\begin{equation}
\label{eq:Lsoft}
  {\cal L}_{soft} = {\cal L}_{soft}^{MSSM} + B_i \epsilon_i {\tilde
    L}_i H_u. 
\end{equation} 
$W_{MSSM}$ is the ordinary superpotential of the MSSM and $i=e,
\mu,\tau$. 
As a result of eq.~(\ref{eq:Lsoft}), 
the scalar neutrinos develop non-zero vacuum expectation $v_i=\langle
{\tilde \nu}_i \rangle$ 
in addition to the VEVs $v_u$ 
and $v_d$ of the MSSM Higgs fields $H_u^0$ and $H_d^0$.  Together with
the bilinear parameters $\epsilon_i$ the $v_i$ induce mixing between
particles distinguished (only) by lepton
number (or R--parity): charged leptons mix with charginos, neutrinos with
neutralinos, and Higgs bosons with sleptons.  
Mixing between the neutrinos and the
neutralinos generates a non-zero
mass for one specific linear superposition of the three neutrino
flavour states of the model at tree-level; the remaining two
masses are generated at 1-loop. 

Charged scalar leptons lighter than all other supersymmetric particles
will decay through R-parity violating couplings. Possible final states
are either $l_j\nu_k$ or $q{\bar q}'$. For right-handed charged
sleptons (${\tilde l}_{Ri}$) the former by far dominate over the
hadronic decay mode.
In the limit $(m_{f_j},m_{\nu_k}) \ll m_{{\tilde f}_i}$ the
two-body decay width  for ${\tilde f}_i \rightarrow f_j +
\Sigma_k \nu_k$ for $i\neq j$ scales as
\begin{equation}\label{width}
\Gamma = \frac{m_{{\tilde f}_i}}{16\pi} 
(\cos\theta_{\ti l_i}Y_{l_i}\frac{\epsilon_j}{\mu})^2,
\end{equation}
which implies that the decay length  $\sim$ Yukawa$^{-2}$.  
The numerical
calculations were performed in the mSUGRA version of the MSSM
by scanning the parameters in the following
ranges: $M_2 \in (0,1.2)$ TeV, $|\mu| \in (0,2.5)$ TeV, $m_0 \in 
(0,0.5)$ TeV, $A_0/m_0$ and $B_0/m_0 \in (-3,3)$ and $\tan\beta
\in (2.5,10)$.  All randomly generated points were subsequently tested for
consistency with the minimization (tadpole) conditions of the Higgs
potential as well as for phenomenological constraints from
supersymmetric particle searches. In addition, points in
which at least one of the charged sleptons was lighter than the
lightest neutralino, and thus the LSP, were selected. This latter 
requirement prefers
strongly $m_0 \ll M_2$. The 
R-parity violating parameters were chosen in such a way 
that the neutrino masses and mixing angles are approximately
consistent with the experimental data.

\begin{figure}[h!]
\setlength{\unitlength}{1mm}
\begin{center}
\epsfig{figure=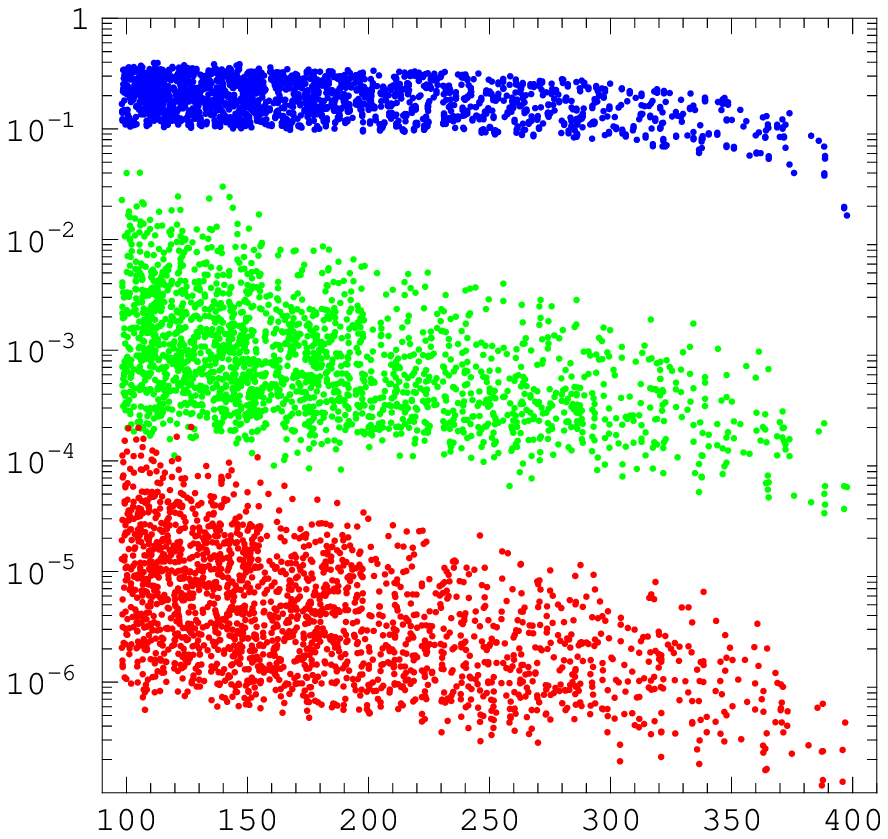,height=5cm,width=5cm}
\hspace{15mm}\epsfig{figure=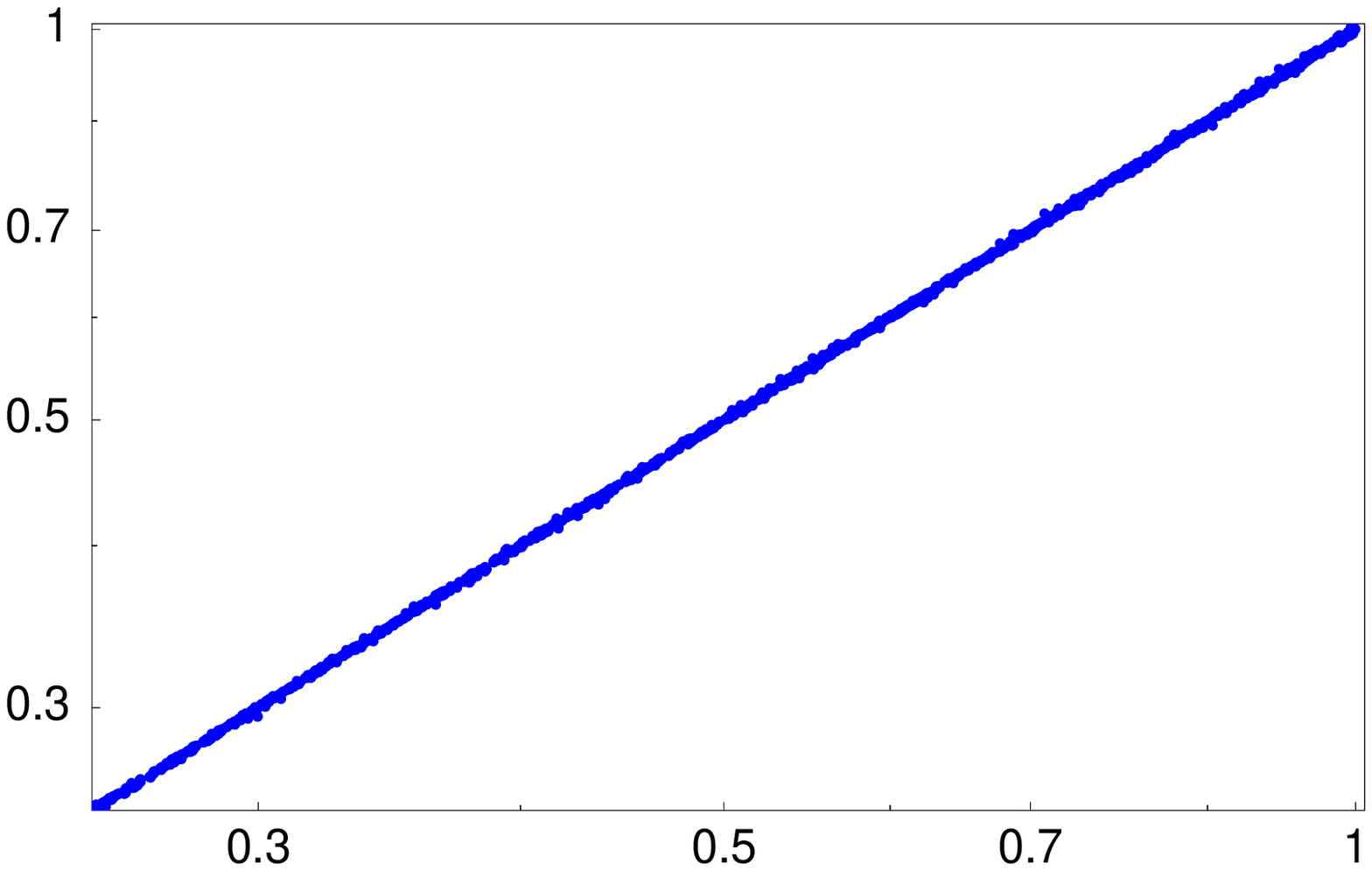,height=5cm,width=5cm}
\put(-110,50){ $L({\tilde l})$~[cm]}
\put(-75,-2){$m_{\tilde l}$ [GeV]}
\put(-53,0){\begin{rotate}{90}
{\small $Br({\tilde \tau}_1 \rightarrow e \sum \nu_i)/
Br({\tilde \tau}_1 \rightarrow \mu \sum \nu_i)$}
\end{rotate}}
\put(-10,-2){$(\epsilon_1/\epsilon_2)^2$}
\end{center}
{Figure 4: Left: Charged slepton decay length 
  as a function of $m_{\tilde l}$ at a linear collider with 0.8 TeV
  c.m.s.  energy. From top to bottom: ${\tilde e}$ (dark, on color
  printers blue), ${\tilde \mu}$ (light shaded, green) and ${\tilde
    \tau}$ (dark shaded, red). Right: Ratios of branching ratios for
scalar tau  decays. }
\label{fig:rpa}
\end{figure}

As seen in Fig.~4 sleptons, including the LSP
$\tilde\tau_R$,  
decay within
the  detector.  The three generations of sleptons decay with quite
different decay lengths and thus it should be possible to separate the
different generations experimentally at a future linear collider.
Note that the ratio of the decay lengths $L(\st)/L(\tm)$ is
approximately given by $(h_{\mu}/h_{\tau})^2$.
Ratios of
branching ratios of various charged slepton decays contain rather
precise information on ratios of the bilinear parameters $\epsilon_i$, 
Fig.~4.\\

\noin {\bf Summary:} Relaxing constraints on the MSSM parameter space
can lead to a variety of striking signals. 
We are still far from understanding all
possible facets of the MSSM, not to mention non-minimal supersymmetric
models. Nevertheless, future $e^+e^-$ colliders will serve as
a powerful tool to unravel the underlying theory.\\

\noin {\bf Acknowledgements:} Work supperted by the European
Community's Human Potential Programme under contracts HPRN-CT-200-00148,
HPRN-CT-2000-00149 and HPMT-2000-00124, the Polish-German LC project
POL 00/015 and the 
Spanish grant BFM2002-00345.  W.P. is
supported by the Erwin Schr\"odinger  fellowship No. J2095 of the 
`Fonds zur F\"orderung der wissenschaftlichen Forschung' of Austria FWF
and partly by the Swiss `Nationalfonds'. M. H.  is supported by a
Spanish MCyT Ramon y Cajal contract.

\end{document}